\documentclass[a4paper,11pt]{article}
\usepackage{pos}
\usepackage{graphicx}

\title{Hybrid Hadronization of Jet Showers from $e^++e^-$ to $A+A$ with JETSCAPE}

\author*[a,b]{Cameron Parker}
\author{the JETSCAPE Collaboration}

\affiliation[a]{Cyclotron Institute, Texas A\&M University,\\
  3366 TAMU, College Station TX 77843, USA}

\affiliation[b]{Department of Physics and Astronomy, Texas A\&M University,\\
  4242 TAMU, College Station TX 77843, USA }

\emailAdd{cameron.parker@tamu.edu}

\abstract{
In this talk we review jet production in a large variety of collision systems using the JETSCAPE event generator and Hybrid Hadronization. Hybrid Hadronization combines quark recombination, applicable when distances between partons in phase space are small, and string fragmentation appropriate for dilute parton systems. It can therefore smoothly describe the transition from very dilute parton systems like $e^++e^-$ to full $A+A$ collisions.
We test this picture by using JETSCAPE to generate jets in various systems. Comparison to experimental data in $e^++e^-$ and $p+p$ collisions allows for a precise tuning of vacuum baseline parameters in JETSCAPE and Hybrid Hadronization. Proceeding to systems with jets embedded in a medium, we study in-medium hadronization for jet showers. We quantify the effects of an ambient medium, focusing in particular on the dependence on the collective flow and size of the medium. Our results clarify the effects we expect from in-medium hadronization of jets on observables like fragmentation functions, hadron chemistry and jet shape.}

\FullConference{HardProbes2023\\
 26-31 March 2023\\
 Aschaffenburg, Germany\\}

\begin{document}
\maketitle

\section{Introduction}
JETSCAPE is a modular, task-based framework for simulating all aspects of heavy-ion collisions \cite{JETSCAPE}. We first concern ourselves with tuning JETSCAPE using a novel hadronization module: Hybrid Hadronization \cite{Han:2016uhh,Fries:2019vws}. This method first uses Monte Carlo recombination \cite{Fries:2003vb,Fries:2003kq} on the partons after the shower stage and then hadronizes the rest with the Lund string model \cite{Andersson:1983ia}. We use Bayesian analysis to tune JETSCAPE with Hybrid Hadronization to CMS and PHENIX data in vacuum proton-proton systems. Although JETSCAPE is
primarily intended to compute heavy-ion collisions, a solid vacuum baseline is needed.

We then examine medium effects on hadronization by modeling how a single jet hadronizes using a brick of quark gluon plasma. Hybrid Hadronization is unique among hadronization models for shower Monte Carlos as it can take into account medium effects on hadronization through recombination of shower partons with thermal partons, and by allowing thermal partons to become part of strings connecting to shower partons.
Both the existence of the medium and flow of the medium are expected to have pronounced effects on the final state hadrons produced. Medium flow both in the direction of the jet and transverse to it should provide flow effects on softer jet hadrons.

\section{Vacuum Systems}
To examine vacuum systems, we model complete proton-proton collisions in JETSCAPE. The initial hard scattering is handled by PYTHIA 8 \cite{Bierlich:2022pfr}. The partons produced are then showered with MATTER until their virtuality is below a cutoff $Q_0$ \cite{MATTER1,MATTER2}. All partons below that cutoff are then hadronized with Hybrid Hadronization. We generate events for $\sqrt{s}=2.76$ TeV and $\sqrt{s}=200$ GeV
to compare to data from LHC and RHIC, respectively. For LHC events we consider data for jets clustered with the
anti-$k_T$ algorithm with various jet radii 
$R$ as well as total charged hadrons at high $p_T$ \cite{CMSJets,CMShadrons}. For RHIC we consider high-$p_T$ pions \cite{PHENIX}. 

We tune 8 different parameters with our setup. In MATTER we adjust the lower virtuality cutoff $Q_0$, the virtuality factor $f=Q_\text{max}^2/p_T^2$ which determines the upper limit for the virtuality of a particle with transverse momentum $p_T$, and $\Lambda_{QCD}$. In the recombination section of Hybrid Hadronization we can rescale the size of the pion, kaon and proton wave functions with parameters called pion width, kaon width, and proton width, respectively. In PYTHIA 8 string
fragmentation we vary the strange to up-down ratio, and the diquark to quark ratio. 

\begin{figure}[h!] \begin{center}
  \includegraphics[width=\textwidth]{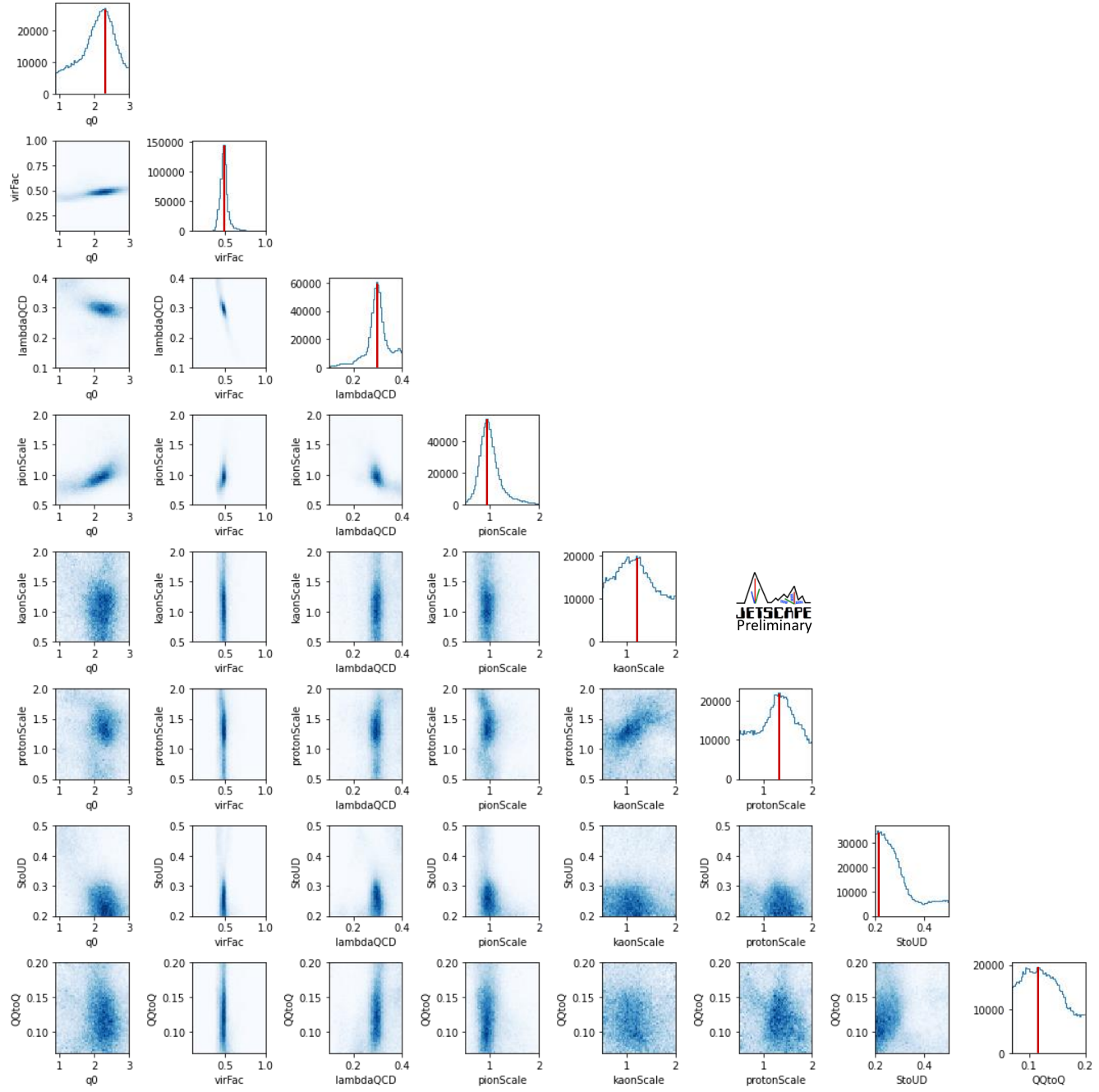}
  \caption{Posterior parameter distributions with highlighted maximum values. Most parameter distributions exhibit solid peaks away from the boundaries of their parameter ranges.}
  \label{fig:params}
\end{center} \end{figure}

\begin{figure}[h!] \begin{center}
  \includegraphics[width=\textwidth]{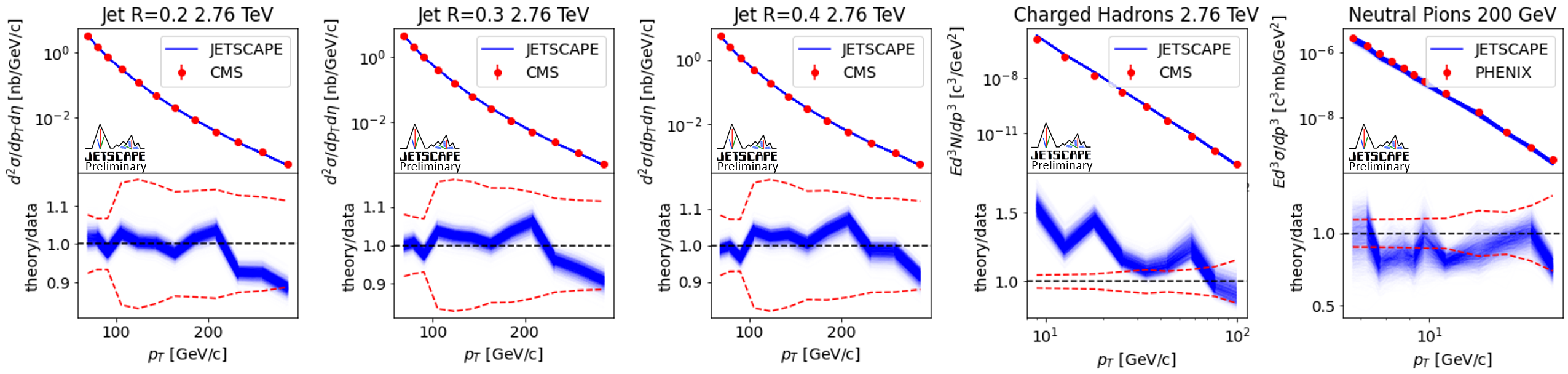}
  \caption{Generated data from the posterior compared to observables. We see a very strong agreement with the jet data and an agreement with hadron data that improves as $p_T$ increases. We compare to CMS data \cite{CMSJets, CMShadrons} for LHC energy and PHENIX data \cite{PHENIX} for RHIC energy.}
  \label{fig:posteriors}
\end{center} \end{figure}

The Bayesian analysis process begins with creating a starting set of design points within a parameter space given by the prior ranges for each parameter. Our prior distribution of parameters is assumed flat within parameters space, and we use a Latin hypercube to generate these points. They will be used to run JETSCAPE. A Gaussian process emulator is utilized to generate observables between the design points in parameter space. The observables produced are then compared to data. A Markov chain Monte Carlo determines new sets of points that improve the description. This process is repeated until convergence, giving us the posterior distribution. The posterior distributions for our set of parameters are shown in Fig. \ref{fig:params} together with correlations between pairs of parameters. The observables for the posterior distribution are shown in Fig. \ref{fig:posteriors}.

The most likely values we find are $Q_0=2.29$ GeV, a virtuality factor $f=0.478$ and $\Lambda_{QCD}=0.292$ GeV
for MATTER. The scale factors for pion, kaon and proton widths are $0.92$, $1.19$, $=1.31$, respectively. The strange to up-down ratio is $0.206$, and the diquark to quark ratio is $0.114$. Note that this preliminary result is for a limited scope of observables. In particular, without identified kaon and proton spectra included the posterior distributions of the last four parameters are rather broad, as expected. We intend to include identified hadrons as well as hadron spectra at
lower $p_T$, for both energies, in the future. We are also building the appropriate event generation infrastructure for $e^++e^-$ collisions and intend to include those observables in the tune as well.

\section{Medium Effects}
In this section we use a simplified event pipeline. Since we are not interested in the entire event, only how the medium affects hadronization, we only examine a single jet in a quark gluon plasma medium. This is done by firing a single parton in the $x$-direction through a medium with a set length and temperatured (a "brick"), showering the parton with MATTER and LBT, and then hadronizing the shower with Hybrid Hadronization. Flow is emulated by adding a set velocity to the thermal partons at the hadronization stage. In the following, longitudinal is defined as in the direction of the jet ($x$) and transverse is perpendicular to the jet.

\begin{figure}[b] \begin{center}
  \includegraphics[width=\textwidth]{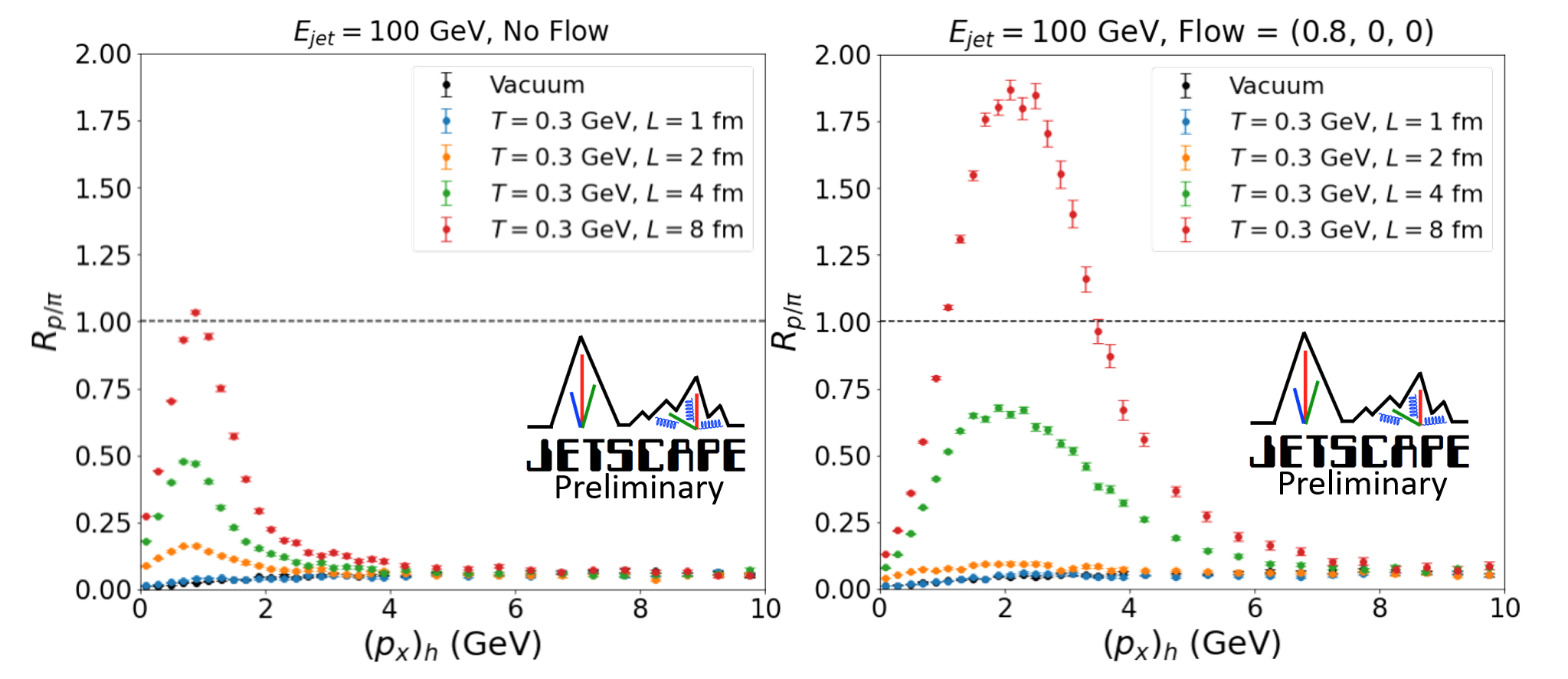}
  \caption{Proton to pion ratios for a jet in bricks of various lengths without flow (left panel) versus with flow (right panel). We see an enhancement in proton production in larger bricks. Flow in the direction of the jet pushes the proton peak to higher $p_T$.}
  \label{fig:ratio}
\end{center} \end{figure}

We first examine the effects of the presence of thermal partons during hadronization. We plot the proton-to-pion ratio as a function of hadron momentum $p_x$ for a variety of different brick lengths from 0 (vacuum) to 8 fm. A large proton-to-pion ratio is known as a signature of quark recombination in hadronization. As shown in the left panel of Fig. \ref{fig:ratio}, the proton-to-pion ratio around 1 GeV indeed increases in magnitude as the brick increases in size. This is consistent with the idea that recombination with thermal partons increases in a larger medium. 

When a homogenous longitudinal flow in jet direction is added,
the peaks in the proton-to-pion ratios grow and are shifted
to $2$-$2.5$ GeV. This can be understood by flowing thermal partons adding more momentum to the hadrons they recombine into. Similar effects can be seen in the $\Lambda$-to-$K$ ratio,
not shown here.

\begin{figure}[t] \begin{center}
  \includegraphics[width=\textwidth]{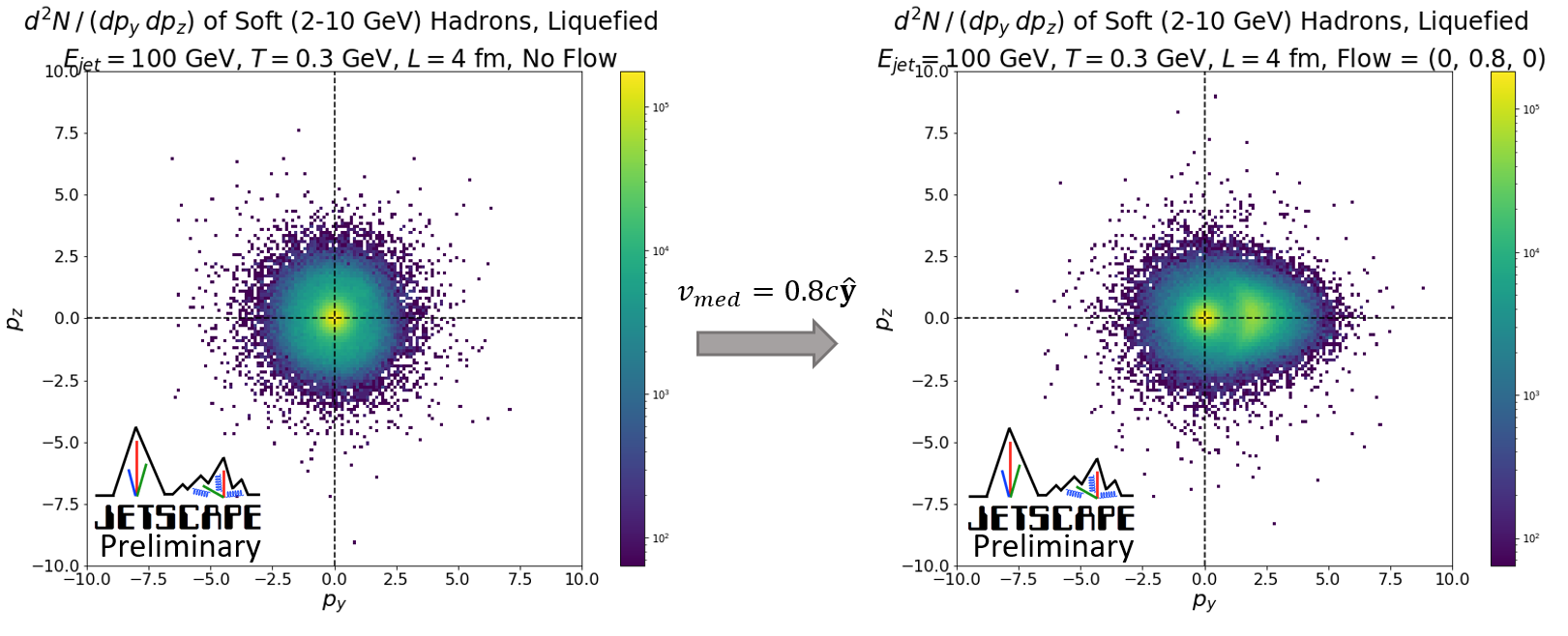}
  \caption{Transverse momentum scatter plot of soft hadrons with no flow (left panel) versus transverse flow (right panel) in the brick. Soft hadrons are noticeably deflected in the direction of the flow due to partons from the medium recombining with shower partons.}
  \label{fig:softhadrons}
\end{center} \end{figure}

We then examine the effects of transverse flow of brick partons. We scatter plot the components of the momentum perpendicular to the jet ($p_y$ and $p_z$) for each hadron to showcase the effects of the flow. Soft hadrons, defined as $2$ GeV $<p_x<10$ GeV, demonstrate a significant deflection in the direction of the flow as shown in Fig. \ref{fig:softhadrons}. On the other hand, leading hadrons demonstrate no noticeable deflection. This is consistent with expectation. Leading partons are distant from thermal partons in phase space, and therefore have a negligible chance to recombine with them.

\section{Conclusion}
We have found a tune for $p+p$ collisions using MATTER
and Hybrid Hadronization which gives acceptable results for
jet and hadron spectra at LHC and RHIC energies, as long
as the hadron transverse momenta are not too small.
 Moving forward we will be incorporating soft hadron spectra, more identified hadron spectra, additional collision energies, and $e^++e^-$ collisions. This will allow us to build a more comprehensive baseline. 
 
 Our study of medium effects in jet hadronization is similarly promising. Using a brick with flow we reproduce all the expected effects, including baryon enhancement increasing with medium size, a shift of the baryon/meson peak in momentum with longitudinal flow, and sideways deflection of soft hadrons with transverse flow. We intend to progress on to heavy flavor jets and simulations of jets in full $A+A$ collisions.

This work was supported by the U.S.\ National Science Foundation under awards 1812431 and 2111568, and under award 2004571 through a subcontract with Wayne State University.

\end{document}